\documentclass[%
 reprint,
superscriptaddress,
 amsmath,amssymb,
aps,
pra,
twocolumn,
]{revtex4-2}

\usepackage[utf8]{inputenc}
\usepackage[english]{babel}
\usepackage{blindtext}
\usepackage{textcomp}
\usepackage{multirow}
\usepackage{graphicx}   
\graphicspath{{figs/}}   
\usepackage{dcolumn}
\usepackage{bm}
\usepackage[hidelinks]{hyperref}
\usepackage{amsmath} 
\usepackage{lipsum} 
\usepackage{xcolor} 
\usepackage{upgreek}
\usepackage{tabularx}
\usepackage{siunitx}    
\usepackage{svg}
\usepackage{svg}
\usepackage{wasysym}
\usepackage{float}
\usepackage{epstopdf}
\usepackage{braket}
\usepackage[title]{appendix}
\usepackage{import}
\usepackage{calc}
\usepackage{natbib}
\usepackage{url}
\usepackage[version=4]{mhchem}
\usepackage{wrapfig}
\DeclareSIUnit\bar{bar}

\begin{document}

\preprint{APS/123-QED}

\title{On-chip stencil lithography for superconducting qubits}

\author{Roudy~Hanna}
 \affiliation{PGI-9, Forschungszentrum J\"ulich and JARA J\"ulich-Aachen Research Alliance, J\"ulich, Germany.}
 \affiliation{RWTH Aachen University, 52062 Aachen, Germany.} 

\author{S\"oren~Ihssen}
 \affiliation{IQMT, Karlsruhe Institute of Technology, 76344 Eggenstein-Leopoldshafen, Germany.}

\author{Simon~Geisert}
 \affiliation{IQMT, Karlsruhe Institute of Technology, 76344 Eggenstein-Leopoldshafen, Germany.}
 
\author{Umut~Kocak}
 \affiliation{PGI-9, Forschungszentrum J\"ulich and JARA J\"ulich-Aachen Research Alliance, J\"ulich, Germany.}
 \affiliation{RWTH Aachen University, 52062 Aachen, Germany.} 
 
\author{Matteo~Arfini}
 \affiliation{Kavli Institute of Nanoscience, Delft University of Technology, Lorentzweg 1, 2628 CJ Delft, The Netherlands.}

\author{Albert~Hertel}
 \altaffiliation[Current address: ]{Qruise GmbH, 66113 Saarbr\"ucken, Germany.}
 \affiliation{PGI-9, Forschungszentrum J\"ulich and JARA J\"ulich-Aachen Research Alliance, J\"ulich, Germany.}

\author{Thomas~J.~Smart}
 \affiliation{PGI-9, Forschungszentrum J\"ulich and JARA J\"ulich-Aachen Research Alliance, J\"ulich, Germany.}
 
\author{Michael~Schleenvoigt}
 \affiliation{PGI-9, Forschungszentrum J\"ulich and JARA J\"ulich-Aachen Research Alliance, J\"ulich, Germany.}

\author{Tobias~Schmitt}
 \affiliation{PGI-9, Forschungszentrum J\"ulich and JARA J\"ulich-Aachen Research Alliance, J\"ulich, Germany.}

\author{Joscha~Domnick}
 \affiliation{PGI-9, Forschungszentrum J\"ulich and JARA J\"ulich-Aachen Research Alliance, J\"ulich, Germany.}

\author{Kaycee~Underwood}
 \affiliation{PGI-9, Forschungszentrum J\"ulich and JARA J\"ulich-Aachen Research Alliance, J\"ulich, Germany.}
 
\author{Abdur~Rehman~Jalil}
 \affiliation{Institute for Experimental Physics III University of Würzburg, 97074 Würzburg, Germany.}
 \affiliation{PGI-10, Forschungszentrum J\"ulich and JARA J\"ulich-Aachen Research Alliance, J\"ulich, Germany.}
 
\author{Jin~Hee~Bae}
 \affiliation{PGI-9, Forschungszentrum J\"ulich and JARA J\"ulich-Aachen Research Alliance, J\"ulich, Germany.}
 
\author{Benjamin~Bennemann}
 \affiliation{PGI-10, Forschungszentrum J\"ulich and JARA J\"ulich-Aachen Research Alliance, J\"ulich, Germany.}

\author{Mathieu~F\'echant}
 \affiliation{IQMT, Karlsruhe Institute of Technology, 76344 Eggenstein-Leopoldshafen, Germany.}

\author{Mitchell~Field}
 \affiliation{IQMT, Karlsruhe Institute of Technology, 76344 Eggenstein-Leopoldshafen, Germany.} 

\author{Martin~Spiecker}
 \affiliation{IQMT, Karlsruhe Institute of Technology, 76344 Eggenstein-Leopoldshafen, Germany.} 

\author{Nicolas~Zapata}
 \affiliation{IQMT, Karlsruhe Institute of Technology, 76344 Eggenstein-Leopoldshafen, Germany.} 

\author{Christian~Dickel}
 \affiliation{Physics Institute II, University of Cologne, 50937 K\"oln, Germany.}

\author{Erwin~Berenschot} 
 \affiliation{Mesoscale Chemical Systems, MESA+Institute, University of Twente, AE Enschede 7500, the Netherlands.}

\author{Niels~Tas} 
 \affiliation{Mesoscale Chemical Systems, MESA+Institute, University of Twente, AE Enschede 7500, the Netherlands.}

\author{Gary~A.~Steele}
 \affiliation{Kavli Institute of Nanoscience, Delft University of Technology, Lorentzweg 1, 2628 CJ Delft, The Netherlands.}

\author{Detlev~Gr\"utzmacher}
 \affiliation{PGI-9, Forschungszentrum J\"ulich and JARA J\"ulich-Aachen Research Alliance, J\"ulich, Germany.}
 \affiliation{RWTH Aachen University, 52062 Aachen, Germany.} 
 \affiliation{PGI-10, Forschungszentrum J\"ulich and JARA J\"ulich-Aachen Research Alliance, J\"ulich, Germany.}
 
\author{Ioan~M.~Pop}
 \affiliation{IQMT, Karlsruhe Institute of Technology, 76344 Eggenstein-Leopoldshafen, Germany.}
 \affiliation{PHI, Karlsruhe Institute of Technology, 76131 Karlsruhe, Germany.} 
 \affiliation{Physics Institute 1, Stuttgart University, 70569 Stuttgart, Germany.}

\author{Peter~Sch\"uffelgen.}
 \email{p.schueffelgen@fz-juelich.de} 
 \affiliation{PGI-9, Forschungszentrum J\"ulich and JARA J\"ulich-Aachen Research Alliance, J\"ulich, Germany.}

                 
\begin{abstract}

Improvements in circuit design and more recently in materials and surface cleaning have contributed to a rapid development of coherent superconducting qubits.
However, organic resists commonly used for shadow evaporation of Josephson junctions (JJs) pose limitations due to residual contamination, poor thermal stability and compatibility under typical surface-cleaning conditions.
To provide an alternative, we developed an inorganic SiO$_2$/Si$_3$N$_4$ on-chip stencil lithography mask for JJ fabrication. 
The stencil mask is resilient to aggressive cleaning agents and it withstands high temperatures up to 1200\textdegree{}C, thereby opening new avenues for JJ material exploration and interface optimization. 
To validate the concept, we performed shadow evaporation of Al-based transmon qubits followed by stencil mask lift-off using vapor hydrofluoric acid, which selectively etches SiO$_2$. 
We demonstrate average $T_1 \approx 75 \pm 11~\SI{}{\micro\second}$ over a 200 MHz frequency range in multiple cool-downs for one device, and $T_1 \approx 44\pm 8~\SI{}{\micro\second}$ for a second device. 
These results confirm the compatibility of stencil lithography with state-of-the-art superconducting quantum devices and motivate further investigations into materials engineering, film deposition and surface cleaning techniques.

\end{abstract}

\maketitle


\section{\label{Sec_Intro}Introduction}

Superconducting quantum circuits have been engineered for a broad range of applications~\cite{braginski2019superconductor}, ranging from photon-~\cite{Peacock1996Single} and particle-detectors~\cite{Day2003Broadband} to the current technological push for quantum computing~\cite{Arute19Quantum,Jurcevic2021Demonstration, acharya2024quantum,Devoret13Superconducting}. In particular, superconducting qubits have steadily improved due to advancements along several axes~\cite{Siddiqi21Engineering,Wallraff04Circuit,Koch07Charge,You07Low-decoherence,Barends13Coherent,Gyenis21Experimental,Long19High}. One of the most important aspects is fabrication process engineering, aiming to understand and reduce the density of two-level systems (TLSs)~\cite{Martinis05Decoherence, Oh06Elimination, Lisenfeld19Electric}, and improving surfaces~\cite{Quintana14Characterization,Wang15Surface,Dial15Bulk,Gambetta16Investigating} and interfaces~\cite{biznarova2024mitigation,Woods19Determining}. This includes the introduction of tantalum (Ta) ground planes and capacitors~\cite{Place21New, bland20252d,Wang22Towards} or the investigation of new capping materials~\cite{Bal24Systematic}. The key non-linear element, the Josephson junction (JJ), has predominantly relied on double-angle evaporation techniques~\cite{Dolan77Offset,Muthusubramanian24Wafer,pishchimova2023improving}. However, there are other approaches being explored to make the process more reproducible and scalable with optical lithography~\cite{vanDamme24Advanced,ke2025scaffold}. In addition to that, recent improvements in gap engineering to suppress quasi-particle tunneling~\cite{Marchegiani2022Quasiparticles,kamenov2023suppression,mcewen2024resisting}, simplified integration with surrounding circuit elements~\cite{Osman21Simplified}, and the exploration of all-nitride electrodes~\cite{Kim21Enhanced}, all contributing to more robust and scalable JJ implementations.


\begin{figure*}[t]
    \includegraphics{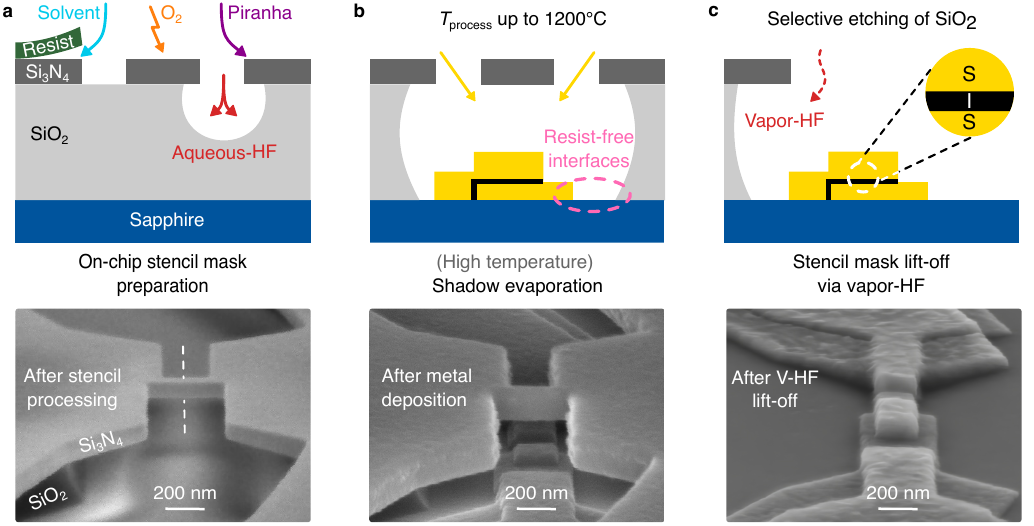}
    \caption{\label{fig1} 
    \textbf{On-chip stencil-lithography fabrication steps of a Josephson junction.}
    The top row schematically presents the stencil fabrication steps while the bottom row shows scanning electron microscopy (SEM) images of a stencil Dolan-bridge~\cite{Dolan77Offset} for a JJ device.
    \textbf{a)} After LPCVD deposition of the inorganic SiO$_2$/Si$_3$N$_4$ bi-layer, the Si$_3$N$_4$ is dry-etched, following a standard e-beam lithography and resist-development, to form the stencil mask. This resist is then cleaned using solvent AR 600-71, an O$_2$ plasma ashing step and a subsequent dip in Piranha solution [H$_2$SO$_4$(96\%):H$_2$O$_2$(31\%) (2:1)]; none of which affect the stencil stack. The stencil mask is then released by selectively etching the SiO$_2$ sacrificial layer against the Si$_3$N$_4$ mask layer using aqueous hydrofluoric acid 1\% (A-HF 1\%). The dotted line in the SEM image indicates the planar cross-section of the corresponding illustrations above.
    \textbf{b)} The junction is fabricated by a standard double-angle evaporation using the shadow stencil mask. In contrast with resist-based methods, the stencil mask can survive temperatures up to 1200\textdegree{}C, allowing surface preparation and annealing in UHV conditions.
    \textbf{c)} Lastly, the stencil mask is lifted-off via an anhydrous vapor-HF (\mbox{V-HF}) process which isotropically and selectively etches SiO$_2$ against the Al-based JJ trilayer (SIS).
    }
\end{figure*} 


Despite significant research and development efforts, resist-based lithography has remained the most commonly used method for fabricating such superconducting quantum devices~\cite{Quintana14Characterization,Lecocq11Junction}.
One of its key advantages is the easy processing through either lift-off ~\cite{Kreikebaum20Improving} or etching ~\cite{Schoof2024Development} of the desired circuit element geometry.
However, in the case of lift-off, the presence of the relatively fragile polymer mask limits pre-growth cleaning methods such as hydrofluoric acid (HF) treatments and deposition temperatures above $\sim$~300\textdegree{}C. This could lead to amorphous or poly-crystalline layers~\cite{Zeng15Direct,Zeng16Atomic}, oxidized substrate surfaces~\cite{Woods19Determining,Gambetta16Investigating} and resist residues (sometimes called `veil of death')~\cite{pop2012fabrication,milchakov2022optimized}.
In the case of etching, better substrate preparation and high-temperature deposition are possible.
However, structuring the layout requires \textit{ex-situ} material-specific and highly selective dry or wet etching~\cite{johnson2019atomic,Wang22Towards,Kim21Enhanced}; which may affect the film or substrate quality, especially in the vicinity of the junction. While using polymer mask has enabled high-performance qubit devices~\cite{Wang22Towards,biznarova2024mitigation,Bal24Systematic, bland20252d}, their shortcomings are becoming increasingly critical towards even higher coherence, calling for new strategies that preserve their respective benefits.

Stencil lithography~\cite{Vazquez-Mena15Resistless} has recently emerged as a resist-free fabrication method, with distinct off- and on-chip implementations. Off-chip approaches decouple substrate preparation from the patterning of free-standing SiN membranes~\cite{Tsioutsios20Free}, aiming to mitigate dielectric loss~\cite{Quintana14Characterization, Gambetta16Investigating}. This decoupling allows for extensive substrate cleaning while avoiding post-processing. However, off-chip methods rely on delicate  lateral spacers for alignment, making them sensitive to tilt between the mask and the wafer. Even small misalignment can lead to imprecise shadowing and blurring effects, ultimately limiting reproducibility and scalability. In addition, the membranes can be affected by tensile strain/stress in the membrane at elevated temperatures ($>$600\textdegree{}C).
In contrast, on-chip stencil lithography integrates the mask directly on the substrate, enabling precise shadowing and compatibility with aggressive cleaning, high temperature processes and ultra-high vacuum (UHV) conditions, depending on the material stack. On-chip masks have been successfully applied to various DC devices~\cite{Welander12Shadow, Hoss99NonOrganic}. However, to our knowledge, on-chip masks have not yet been applied to the fabrication of coherent superconducting quantum devices. This is likely due to the robustness of the mask, which complicates lift-off after junction fabrication, and the potential performance limitations introduced by lossy dielectrics in high-frequency circuits if the stencil mask is not lifted off.

In this work, we develop an on-chip stencil lithography technique based on a pre-patterned silicon oxide/silicon nitride (SiO$_2$/Si$_3$N$_4$) inorganic mask ~\cite{Schüffelgen19Selective, Schmitt22Integration}, with the goal of making it compatible with the fabrication of coherent superconducting qubits.
Our stencil mask enables aggressive cleaning chemicals and has been tested to withstand high-temperature annealing up to 1200\textdegree{}C in UHV conditions (see Fig.~\ref{figA1} in App.~\ref{App_Methods}), making it well suited for future material exploration and interface investigation. The crucial chip-wide lift-off of the stencil mask is achieved by selectively etching the SiO$_2$ with vapor hydrofluoric acid (\mbox{V-HF}) through openings in the Si$_3$N$_4$ layer, without attacking the deposited materials.
As a validation of the concept we show data on the coherence of two frequency-tunable Al-based transmon qubits fabricated with this approach.

The paper is structured as follows: Sec.~\ref{Sec_Fab} presents the developed stencil lithography technique. Sec.~\ref{Sec_Device} shows the characterization of Al-based tunable transmon qubits fabricated via the stencil mask. 
Lastly in Sec.~\ref{Sec_Outlook}, we provide a conclusion and discuss possible applications for this technology.


\section{\label{Sec_Fab}Stencil Fabrication}

To achieve this, we developed an inorganic on-chip mask fabrication to pre-pattern our devices. 
We initially deposit two layers via low-pressure chemical vapor deposition (LPCVD): a~\SI{300}{\nano\meter} SiO$_2$ used as a sacrificial layer and a~\SI{100}{\nano\meter} Si$_3$N$_4$ used as a mask layer.
Both layers are deposited on a c-plane sapphire (Al$_2$O$_3$) wafer following HNO$_3$ cleaning, see App.~\ref{App_Methods}. This specific material combination enables two key fabrication steps: SiO$_2$ can be selectively etched over Si$_3$N$_4$ using aqueous hydrofluoric acid (A-HF 1\%) prior to deposition, and over Al/AlO$_\mathrm{x}$ for mask lift-off via vapor-HF after deposition. 

As shown in Fig.~\ref{fig1}a, the Si$_3$N$_4$ is patterned using a standard, resist-based, electron beam (e-beam) lithography technique. To shape the mask layer, the top Si$_3$N$_4$ layer is anisotropically etched via reactive ion etching (RIE) using a CHF$_3$:O$_2$ gas mixture. Afterward, the e-beam resist is dissolved in AR 600-71 solvent, followed by an additional O$_2$ plasma ashing step.

At this stage, the wafer is diced into the desired sample size after applying a protective dicing resist. The diced samples are cleaned and exposed to another O$_2$ plasma ashing step. It is worth noting that, although polymer resists are used in these steps, they do not come into contact with the sapphire surface, as the sacrificial layer is not removed yet. 
To further clean any organic leftovers on the mask, the samples are dipped in a Piranha solution [H$_2$SO$_4$(96\%):H$_2$O$_2$(31\%) (2:1)]; which does not attack neither the SiO$_2$ nor Si$_3$N$_4$ layers. 

\begin{figure}[b]
    \includegraphics{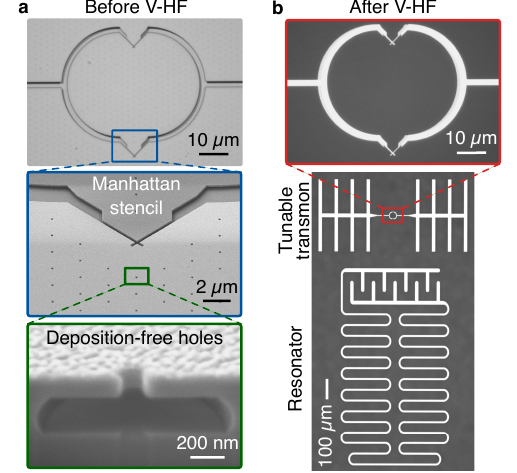}
    \caption{\label{fig2} 
    \textbf{Stencil fabrication of an Al-AlO$_\mathrm{x}$-Al transmon qubit.}
    \textbf{a)} The tunable transmon consists of two superconducting islands connected via a superconducting loop interrupted by two JJs, forming a SQUID device. Zoom-in (blue): The design features Manhattan-style junctions and a hexagonal grid of holes for faster \mbox{V-HF} lift-off, see App.~\ref{figA3}. Their size is optimized to block unwanted deposition while allowing the \mbox{V-HF} gas to penetrate through. Zoom-in (green): A focused-ion-beam (FIB) cut showing the deposition-free substrate underneath one of the holes after Al-evaporation. See App.~\ref{App_Methods} for more details.
    \textbf{b)} Gray-scaled microscope images of the entire transmon qubit circuit after \mbox{V-HF} liftoff. The white areas correspond to the deposited Al-trilayer while the black area is the sapphire substrate. Zoom-in (red): The SQUID loop with both JJs and the superconducting ring after mask removal.
    }
\end{figure}

From this point onward, our technique ensures a \mbox{\textit{resistless}}, single-step \textit{in-situ} stencil lithography.
To expose the surface of the substrate and release the mask, the SiO$_2$ is isotropically and selectively etched against the Si$_3$N$_4$ via A-HF 1\%, see Fig.~\ref{fig1}a. This creates an under-etch region around the edges of the Si$_3$N$_4$ mask structures, thus preventing sidewall fencing during deposition and lift-off procedures. See App.~\ref{App_Methods} for more information. We have tested the stability of the inorganic stencil mask up to 1200\textdegree{}C. This high temperature resilience can be leveraged in future experiments to add another surface treatment prior to deposition (see App.~\ref{App_Methods}). Once the stencil mask is prepared, we perform standard double-angle shadow e-beam evaporation to fabricate transmon qubits with Al superconducting electrodes (S) and Al/AlO$_\mathrm{x}$/Al junctions, where the in situ formed, non-stoichiometric AlO$_\mathrm{x}$ serves as the insulating layer (I). The stencil mask is shown to be compatible with both Dolan (Fig.~\ref{fig1}) and Manhattan-style junction layouts (Fig.~\ref{fig2}).

Lastly, for superconducting qubit applications, in contrast to DC devices, it is crucial to lift-off the stencil mask~\cite{Welander12Shadow} to eliminate the dielectric losses and parasitic capacitive coupling induced by having the inorganic stack and the metal on top of it.
Therefore, the mask is removed by selectively etching the SiO$_2$ via \mbox{V-HF}~\cite{Witvrouw00Comparison} against the now functional S-I-S layers; shown in Fig.~\ref{fig1}c. This method has been previously shown to have a minimal impact on the performance of Al resonators~\cite{Dunsworth18Method} and has more recently been explored for scaffolding-assisted junction fabrication~\cite{ke2025scaffold}, as AlO$_\mathrm{x}$ acts as an effective etch-stop material for \mbox{V-HF}~\cite{Bakke05Etch}. The sample is finally dipped in de-ionized (DI) water followed by isopropanol (IPA) to rinse off etching residues from the lift-off.
Scanning electron microscopy (SEM) images of a stencil-made JJ before and after mask removal are also presented in Fig.~\ref{fig1}.
More details can be found in App.~\ref{App_Methods}.


\begin{figure*}[t]
    \includegraphics{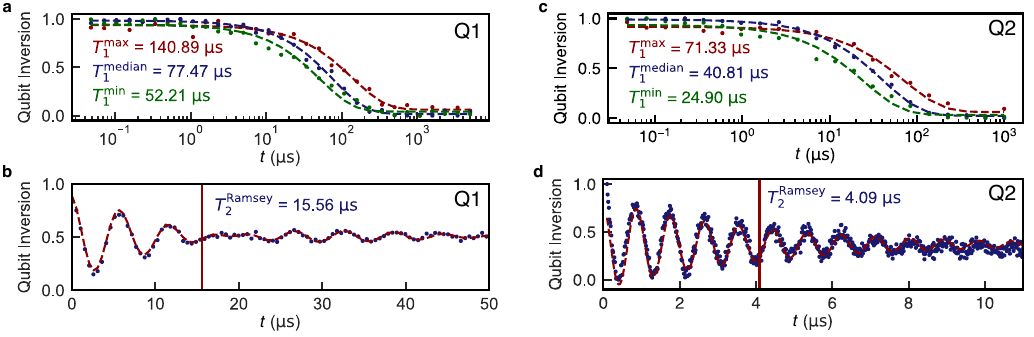
    }
    \caption{\label{fig3}
    \textbf{Time-domain measurement of the stencil transmon qubits Q1 and Q2.}
    \textbf{a,c)} Decay curves of maximal, median and minimal qubit $T_1$ lifetimes (red, blue and green, respectively) within a measurement period of 60 hours and 3.5 hours for Q1 and Q2, respectively. The measurements are taken at zero external flux bias.
    \textbf{b,d)} $T_2^\mathrm{Ramsey}$ coherence measurement where a beating pattern is visible due to a \SI{0.028}{\mega\hertz} and \SI{2.5}{\mega\hertz} charge dispersion for each qubit respectively. More measurements are provided in App.~\ref{App_Meas}.
    }
\end{figure*}


\section{\label{Sec_Device} Stencil Qubit}

To test the validity of the on-chip stencil lithography fabrication, we use a tunable transmon qubit layout, as shown in Fig.~\ref{fig2}a,b. 
The design features a lumped element resonator capacitively coupled to two islands connected through a superconducting quantum interference device (SQUID) with two nominally identical Manhattan-style Al-AlO$_\mathrm{x}$-Al junctions. The chip contains 4 resonator-qubit pairs and one test resonator.
To enable the \mbox{V-HF} lift-off, we integrate a hexagonal grid of holes across the sample. Their size is optimized to block unwanted deposition while allowing isotropic etching via \mbox{V-HF} as explained in App.~\ref{App_Methods}.

After \mbox{V-HF} exposure and mask lift-off, we characterize the stencil-based qubits via standard circuit quantum electrodynamics measurements in the dispersive readout regime~\cite{Blais2021CQED}. 
The sample is mounted in a 3-dimensional copper waveguide (as in Fig.~\ref{figA7}) with a~\SI{6}{\giga\hertz} cutoff frequency, similar to Ref~\cite{Kou2018Simultaneous, Grunhaupt19Granular}. The waveguide is equipped with a global flux tuning bias coil, a qubit drive port, and a readout port through which the sample is measured in reflection via a circulator. 
After assembly, the waveguide is placed inside an aluminium and mu-metal magnetic shield and attached to the mixing chamber stage of the cryostat for thermalisation to base temperature ($\approx$~\SI{20}{\milli\kelvin}). See App.~\ref{App_Meas} for more details on the cryogenic setup. 

\begin{table}[b]
\centering
\setlength{\tabcolsep}{12pt} 
\begin{tabular}{lcc}
\textbf{Parameter} & \textbf{Q1} & \textbf{Q2} \\
\hline
$\omega_{\mathrm{01}}/2\pi$ (GHz) & 3.112 & 3.480 \\
$\alpha/2\pi$ (MHz)               & -202  & -350 \\
$f_\mathrm{r}$ (GHz)                       & 7.323 & 6.471 \\
$\kappa/2\pi$ (MHz)               & 0.223 & 0.544 \\
$\chi/2\pi$ (MHz)                 & -0.35  & -1.5 \\
$E_\mathrm{J}/2\pi$ (GHz)                  & 7.81  & 6.45 \\
$E_\mathrm{C}/2\pi$ (GHz)                  & 0.17  & 0.28 \\
SQUID asymmetry (max.)            & 5\%   & -- \\
\hline
\end{tabular}
\caption{\textbf{Extracted parameters for Q1 and Q2:} transition frequency $\omega_{\mathrm{01}}$, anharmonicity $\alpha$, resonator frequency $f_\mathrm{r}$, resonator linewidth $\kappa$, dispersive shift $\chi$, Josephson energy $E_\mathrm{J}$, charging energy $E_\mathrm{C}$, and estimated SQUID asymmetry.}
\label{tab:qubit_params}
\end{table}

We report here the measurement results on 2 qubits on the same chip, Q1 and Q2, summarized in Table~\ref{tab:qubit_params}. The two additional qubits on the chip were not operational due to suspected electrostatic discharge during handling or measurement.
In Fig.~\ref{fig3} we show the measured $T_1$ energy relaxation times for the stencil qubits. For Q1, we observe an average energy relaxation time of $\overline{T_1} \approx 75.37 \pm 7.42$~\SI{}{\micro\second} over a 60-hour measurement period (see App.~\ref{App_Meas}). The exponential decay of the maximum (red), median (blue), and minimum (green) lifetimes are measured to be~\SI{140}{\micro\second},~\SI{77}{\micro\second}, and~\SI{52}{\micro\second}, respectively. The mean value corresponds to a qubit quality factor of $\sim 1.5\times10^6$.
The Ramsey dephasing time is measured to be $T_2^\mathrm{Ramsey} \approx$~\SI{15}{\micro\second}, and a $T_2^\mathrm{Echo} \approx$~\SI{28}{\micro\second}. 
Similar experiments were also performed on Q2 which exhibits $\overline{T_1} \approx 44.28 \pm 8.23$~\SI{}{\micro\second} and $T_2^\mathrm{Ramsey} \approx$ 4~\SI{}{\micro\second}. The low $T_2^\mathrm{Ramsey}$ values could be attributed to charge noise of the qubits. Complementary measurements are provided in App.~\ref{App_Meas}.

\begin{figure}[t]
    \includegraphics{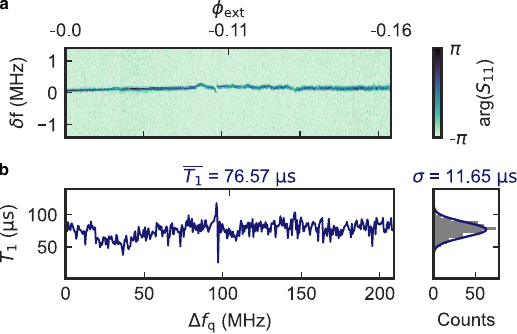}
    \caption{\label{fig4}
    \textbf{Two-tone spectroscopy over flux of Q1.}
    \textbf{a)} Two-tone spectroscopy of the transmon qubit while tuning its frequency ($\text{f}_\mathrm{q}(\phi_{ext}=0)=3.114$ GHz) with an external flux ($\phi_{ext}$) over~\SI{200}{\mega\hertz}. The qubit frequency is continuously tracked with a~\SI{3}{\mega\hertz} span range. The characteristic parabolic curve is here converted to a linear one for better visibility: The x-axis shows the expected qubit frequency $\Delta \text{f}_\mathrm{q}$ from the circuit model, defined such that $\Delta \text{f}_\mathrm{q} = 0$ corresponds to \SI{3.112}{\giga\hertz} and the y-axis is its deviation from the measured spectrum $\delta$f.
    \textbf{b)} For every flux-point, we perform a $T_1$ measurement and extract its value. The mean $T_1$ is around 76.57$\pm$11.65~\SI{}{\micro\second} over the same frequency range. 
    }
\end{figure}

In the following, we focus the discussion on Q1 to study its flux and time stability in more details. We apply an external flux to measure its response away from the zero-flux sweet spot. We track the 01 transition of Q1 across~\SI{200}{\mega\hertz} using two-tone spectroscopy (see Fig.~\ref{fig4}a). Within this range, we observe two avoided level crossings on the order of a few~\SI{}{\kilo\hertz}. At each flux point, we measure the qubit's relaxation time $T_1$, as shown in Fig.~\ref{fig4}b. The measured $T_1$ values remain relatively stable across the frequency range, similar to Ref.~\cite{thorbeck2023two}, indicating that our stencil method is compatible with state-of-the-art surface-engineered superconducting qubits. The average relaxation time across all flux points is $\overline{T_1} = 76.57 \pm 11.65~\SI{}{\micro\second}$.

\begin{figure}[t]
    \includegraphics{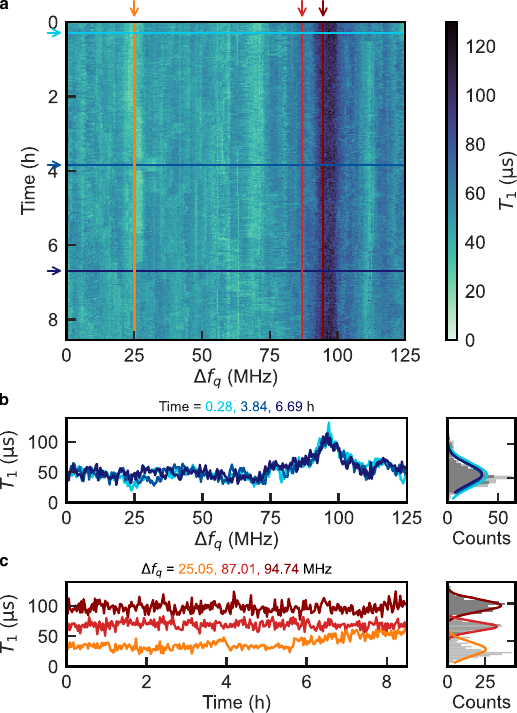}
    \caption{\label{fig5}
    \textbf{Spectral and time resolved coherence measurements of Q1.}
    \textbf{a)} Spectral and temporal resolution of $T_1$ in a different cool-down than before. Every point in this plot represents the lifetime of a decay curve measured with 50 stroboscopic projective qubit measurements spaced~\SI{10}{\micro\second} apart.
    \textbf{b-c)} Line-cuts of $T_1$ as a function of flux and time, shown for three representative points each with the corresponding distributions shown in right panels. The sigma values are reported in App.\ref{App_Meas}. 
    }
\end{figure}

In a different cool-down of the same chip, we spectrally and temporally resolve $T_1$ over a period of 8 hours and a~\SI{125}{\mega\hertz} frequency range in Fig.~\ref{fig5}. Each decay time is extracted by averaging over sequences of 50 stroboscopic qubit measurements, separated by~\SI{10}{\micro\second}, after preparing the qubit in the excited state with an initial $\pi$-pulse. As shown in the line-cuts in Fig.~\ref{fig5}b, these values remain rather stable over time as a function of flux. However, we notice that $T_1$ fluctuates by almost an order of magnitude versus flux in Fig.~\ref{fig5}c, yet it consistently remains above~$\approx$~\SI{20}{\micro\second}.

A detailed analysis of the energy relaxation mechanisms is beyond the scope of this work and will be investigated in future studies. Possible explanation include defects within the junction materials \cite{Zeng15Direct}, short-term fluctuations in the TLS environment \cite{thorbeck2023two,spiecker2023two}, non-equilibrium quasi-particle poisoning \cite{Martinis09Energy,Pop14Coherent}, or other loss mechanisms.


\section{\label{Sec_Outlook}Conclusion \& Outlook}

To summarize, we implemented on-chip stencil lithography based on a SiO$_2$/Si$_3$N$_4$ material stack to fabricate and investigate Al-based coherent transmon qubits. 
The combination of this inorganic stack, together with aggressive cleaning like Piranha and aqueous-HF solutions, ensures resist-free substrate interface prior to material deposition. 
The mask can be lifted-off via vapor-HF through the integrated grid of holes, without etching the functional Al-trilayer. We measured an average lifetime of a transmon qubit of $T_1 \approx $~\SI{76}{\micro\second} and a coherence time $T_2^\mathrm{Ramsey} \approx $~\SI{15}{\micro\second} which are similar to conventional transmons with resist-based fabrication~\cite{Osman21Simplified}. Furthermore, we measured the spectral purity and the stability of $T_1$ over time and frequency, which showed similar characteristics as reported in the literature~\cite{klimov2018fluctuations, thorbeck2023two}. To conclude, we demonstrated that replacing organic resist with our inorganic mask does not compromise the functionality of standard S-I-S transmons, even without exploiting the thermal stability of the stencil. The presented technique opens the way for the exploration of new junction materials and surface cleaning methods. By enabling high-temperature processing and aggressive cleaning steps, it may play an instrumental role in overcoming the decoherence bottleneck currently limiting superconducting qubit technology. In future work, we aim to leverage the stencil’s thermal and chemical robustness to further push the limits of qubit performance.


\begin{acknowledgments}
We are grateful for Rami Barends and Pavel Bushev for many fruitful discussions. We also thank Yebin Liu, Yorgo Haddad, Dmitriy Volkov, Anne Schmidt and Josua Thieme for their valuable input. 

We acknowledge the Helmholtz Nano Facility (HNF) cleanroom staff at the Forschungszentrum J\"ulich (FZJ) for providing the possibility to develop this process. In particular Hubert Stumpf, Thomas Grap, Georg Mathey, Anja Zaß, Christoph Krause, Stephany Bunte and Elmar Neumann for their involvement in various bits of the fabrication and its analysis. Florian Lentz and Stefan Trellenkamp are additionally thanked for performing electron-beam lithography. We appreciate the creative solutions of \mbox{Johannes} Pfennings at the FZJ workshop for designing custom pieces for our setup. We acknowledge the technical support of Patrice Brenner for the Focused Ion Beam (FIB) and Scanning Electron Microscope (SEM) imaging at the KIT Nanostructure Service Laboratory. Similarly, we are thankful for Hande Aydogmus at EKL Lab cleanroom in TU Delft for sharing her expertise using the uEtch. We also thank KLA corporation (formerly SPTS) for helpful vapor-HF discussions. We acknowledge the measurement software framework qKit.

This work was financed by the German Federal Ministry of Research, Technology and Space (BMFTR) within the following projects: QSolid (FKZ:13N16151 and 13N16149), TLE4HSQ (Grant No.13N15983) and Quantum Future project “MajoranaChips” (Grant No.13N15264). We acknowledge the support from the Deutsche Forschungsgemeinschaft (DFG, German Research Foundation) under Germany’s Excellence Strategy – Cluster of Excellence Matter and Light for Quantum Computing (ML4Q) EXC 2004/1 – 39053469.

M.Fi. and N.Z.~acknowledge funding from the European Union under the Horizon Europe Program, grant agreement number 101080152 (TruePA).
M.Sp.~acknowledges partial funding from the German Federal Ministry of Research, Technology and Space (BMFTR) within the project GEQCOS (FKZ:~13N15683).

M.A. and G.A.S. acknowledge support by the Dutch Research Council (NWO) under the project number VI.C.212.087 of the research program VICI round 2021.

\end{acknowledgments}

\section*{Competing Interests}

The authors declare no competing interests.

\section*{Data Availability}

Raw data as well as all measurement, data-analysis, and simulation code used in the generation of main and supplementary figures are available in Zenodo with the identifier: 10.5281/zenodo.15976967.

\newpage

\section*{Author Contributions}

\textbf{R.H.}~developed the stencil process on sapphire, integrated it into superconducting qubit fabrication and wrote the original manuscript with feedback from all other co-authors.

\textbf{S.I.}~designed and simulated the transmon qubit and performed aluminum evaporation. 

\textbf{S.G.}~carried out aluminum evaporation and conducted the qubit measurements. 

\textbf{U.K.}~characterized stencil fabrication and optimized the hole geometry. 

\textbf{M.A.}~and \textbf{G.A.S.}~performed mask lift-off using the uEtch tool in Delft. 

\textbf{A.H.}~assisted with data analysis. 

\textbf{T.J.S.},~\textbf{M.Sc.},~and \textbf{B.B.}~conducted TLE annealing, EDX, and AFM measurements. 

\textbf{T.S.}~and \textbf{J.D.}~contributed to fabrication process development. 

\textbf{K.U.}~performed AFM characterization. 

\textbf{J.H.B.}~conducted FIB and TEM measurements. 

\textbf{A.R.J.}~supported TEM analysis. 

\textbf{M.Fi.}~and \textbf{N.Z.}~contributed to qubit measurements. 

\textbf{M.Fé.}~and \textbf{M.Sp.}~contributed to qubit design and measurements. 

\textbf{C.D.}~performed evaporation tests and supported measurement analysis. 

\textbf{E.B.}~and \textbf{N.T.}~optimized the LPCVD-related processes.

\textbf{D.G.},~\textbf{I.M.P.},~and \textbf{P.S.}~conceived and supervised the project.


\appendix

\setcounter{figure}{0}  
\renewcommand{\thefigure}{\Alph{section}\arabic{figure}}  

\newpage
\onecolumngrid

\section{\label{App_Methods}Methods}

\paragraph{Stencil Fabrication}

\begin{figure*}[h]
    \includegraphics{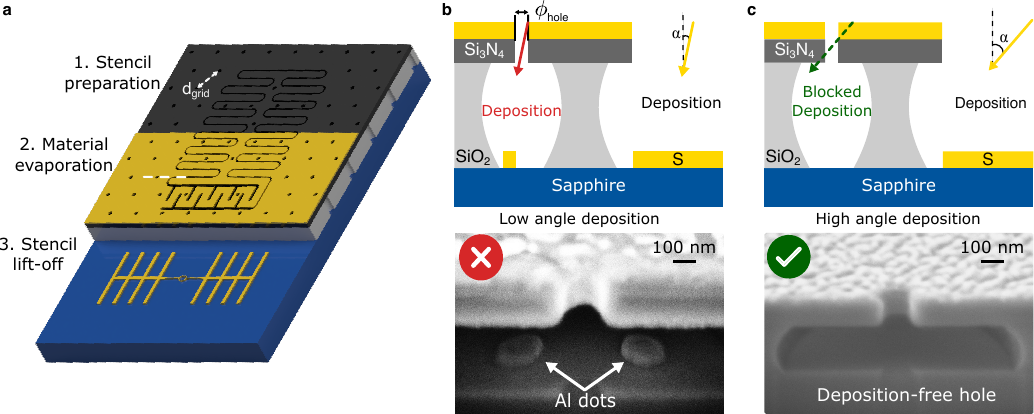}
    \caption{\label{figA1}
    \textbf{On-chip stencil fabrication details.}
    \textbf{a)} Three main steps of the stencil transmon qubit fabrication: 1) stencil mask preparation and integration of a hexagonal grid of holes, 2) double angle shadow-evaporation of Al, and 3) stencil mask lift-off via \mbox{V-HF}.
    The white dotted line represents the cross section for the following images.
    \textbf{b)} Schematic and FIB cut image of a test deposition of material under low angle (i.e. $\alpha$ = 40\textdegree{}) showing double-deposition through the hole.
    \textbf{c)} Optimized deposition under a high angle (i.e. $\alpha$ = 70\textdegree{}) where the hole blocks the material from reaching the substrate, see paragraph \textit{d}.
    }
\end{figure*}

The on-chip stencil mask lithography is done on wafer scale using single-side polished 2" HEM\textsuperscript{\textregistered} sapphire from Crystal Systems LLC \cite{Crystal-Systems-LLCHEM}.
The wafers are first cleaned in 100\% HNO$_{3}$ twice, each for 5 min, followed by a final 10 min dip in 69\% HNO$_{3}$. 
After a DI-water rinse, the wafers are transferred to the low-pressure chemical vapor deposition (LPCVD) chamber (Tempress Systems BV).
First, a~\SI{300}{\nano\meter} layer of silicon oxide (SiO$_2$) is grown homogeneously by Tetraethylorthosilicate (Si(OC$_2$H$_5$)$_4$ or TEOS) evaporation at $T$ = 725\textdegree{}C, $p$ = 200 mTorr and $Flow_{\mathrm{TEOS}}$ = 40 sccm. It acts as the sacrificial layer throughout the process.
Afterward, a~\SI{100}{\nano\meter} stoichiometric silicon nitride (Si$_3$N$_4$) film is deposited on top of the SiO$_2$, forming the mask layer. This process is done at  $T$ = 800\textdegree{}C, $p$ = 200 mTorr, $Flow_{\mathrm{Si_2Cl_2}}$ = 22 sccm and $Flow\mathrm{_{NH_3}}$ = 66 sccm~\cite{martinussen2024thick}. 
The wafer is then cleaned with Acetone (Ace) and Isopropanol (IPA) and then baked at 110\textdegree{}C for 5 min to remove surface moisture. CSAR 62~\cite{CSAR62.09} e-beam (positive) resist is spin coated at the top of the stencil stack for patterning. For sapphire wafers, an extra (water soluble) conductive resist (Electra 92~\cite{Electra92} or E-spacer 300z~\cite{Espacer300z}) is used to prevent charging effects.
After e-beam exposure, the conductive resist is removed in DI-Water. Next, we perform cold development of the CSAR 62 in AR 600-546 developer set at -1\textdegree{}C. The wafer is continuously rotated back and forth between clockwise and counter-clockwise directions for~\SI{70}{\second}. The developed parts now expose the Si$_3$N$_4$ that will define the stencil structures. 
We use CHF$_3$:O$_2$ (55:5\,sccm) gas mixture to anisotropically dry-etch the mask layer with vertical sidewalls in the reactive ion etcher (RIE), see Fig.~\ref{fig1} in the main text and Fig.~\ref{figA1}a.
Because of the homogeneity of the LPCVD, the Si$_3$N$_4$ on the backside of the wafer is also etched using the same recipe by flipping the wafer upside down. After that, the CSAR 62 is removed by an overnight dip in the AR600-71 solvent. An O$_2$ plasma ashing (600\,sccm, 600\,W, 5\,min) step is also done to further clean organic residues.
The processed wafer is then sent to dicing after spin coating it with MC-PC 20 protective resist. This resist on the individual chips is dissolved in Acetone and IPA followed by another O$_2$ plasma ashing exposure.
To further ensure that no organic residues reach the surface and thanks to the inorganic properties of the stencil mask, a 10 min dip in Piranha solution [H$_2$SO$_4$(96\%):H$_2$O$_2$(31\%) (2:1)] is performed. 
Following that, the~\SI{300}{\nano\meter} SiO$_2$ is selectively etched against the Si$_3$N$_4$ by aqueous hydrofluoric acid 1\% (A-HF) solution for 18 min. This releases the stencil mask and effectively also creates an under-etch region of $UE_{\mathrm{A-HF}} \sim$~\SI{450}{\nano\meter}, see Fig.~\ref{figA1} and~\ref{figA3}.

\begin{figure*}[h]
    \includegraphics{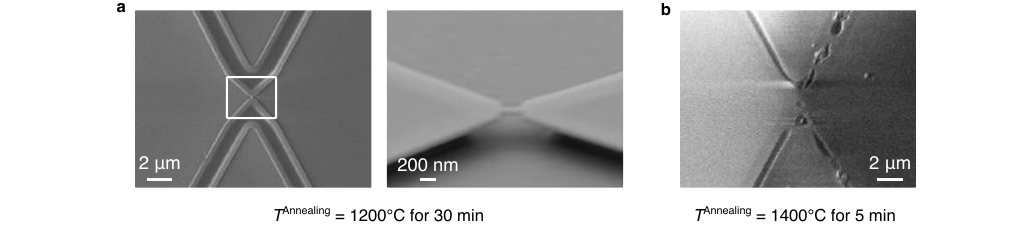}
    \caption{\label{figA5}
    \textbf{SEM images showing the thermal stability of the stencil mask via CO$_2$ laser heating in vacuum.}
    \textbf{a)} The stencil mask remains stable after annealing at $T = 1200$\textdegree{}C for 30 min. 
    The free-standing bridge is visible in the zoomed-in area indicated by the white box.
    \textbf{b)} If we anneal our substrate at higher temperatures, in this case $T = 1400$\textdegree{}C, the stencil masks breaks down and melts away.
    }
\end{figure*}

\paragraph{Thermal Stability of the stencil mask}
To test the thermal stability of the stencil mask, we heat a structured sample in ultra-high vacuum condition via a $\lambda =$10\,~\SI{}{\micro\meter} CO$_2$ substrate laser heater which forms part of our Thermal Laser Epitaxy (TLE) chamber~\cite{Braun20In}.
Using this substrate heater, a large temperature window up to 2000\textdegree{}C is accessible in UHV conditions~\cite{Smart24Twin, Kim23Carbon}.
As shown in Fig.~\ref{figA5}, we observe that the stencil mask remains intact up to 1200\textdegree{}C over an annealing time of 30 minutes, meaning it is highly applicable for the growth of most superconducting materials of interest. Beyond 1200\textdegree{}C, free-standing bridge structures begin to buckle and become structurally compromised due to the desorption of SiO$_2$ from the sapphire substrate. The destruction of the mask at an anneal temperature of 1400\textdegree{}C is shown in Fig.~\ref{figA5}b.

\paragraph{Material Evaporation} 

Both Dolan-style and Manhattan-style junction were successfully fabricated with the on-chip stencil lithography, see Fig.\ref{fig1},\ref{fig2} in main the text. For the qubit device, we chose Manhattan junction to achieve the desired overlap area considering the thickness of our stencil stack and deposition angle without unintended deposition through the holes (see paragraph \textit{d} below).

\begin{figure*}[h]
    \includegraphics{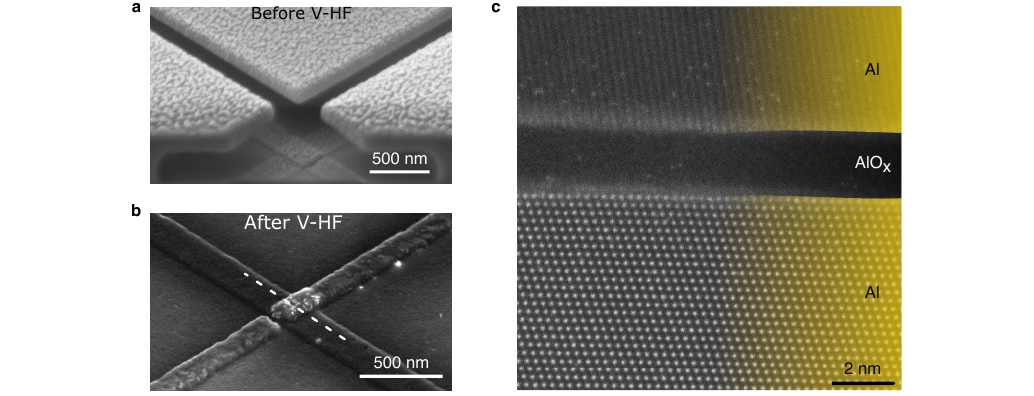
    }
    \caption{\label{figA2}
    \textbf{Manhattan-style JJ and TEM of its trilayer.}
    \textbf{a-b)} SEM images of before and after \mbox{V-HF} stencil mask lift-off of two different samples. For panel a) a FIB is used to take away parts of the mask and expose the junction after deposition for visibility. Panel b) shows a JJ after mask lift-off and before the FIB-cut is performed parallel to the dashed line to create a thin-lamella for analysis the next panel. 
    \textbf{c)} A cross-section of the Al-AlO$_\mathrm{x}$-Al tri-layer lamella taken via a high-angle annular dark-field scanning transmission electron microscopy (HAADF-STEM) image. The beam line is focused along the zone axis of the bottom Al electrode. This shows the sharp S-I interface between the bottom crystalline superconductor and the amorphous insulator. The tunnel barrier thickness is approximately~\SI{2}{\nano\meter}. The top electrode also shows a crystalline structure but appears blurry due to a rotation misorientation relative to the focused bottom Al zone axis. 
    }
\end{figure*}

\begin{figure*}[b]
    \includegraphics{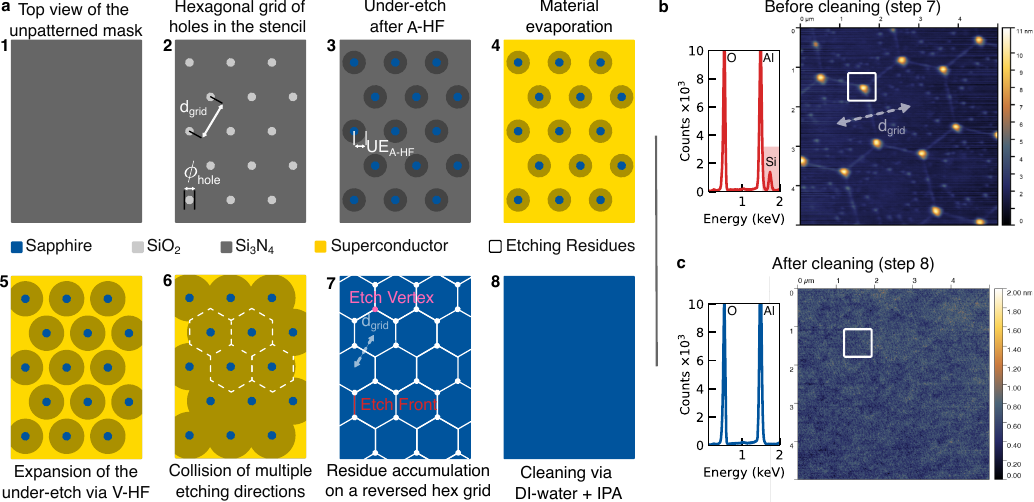
    }
    \caption{\label{figA3}
    \textbf{Stencil mask lift-off via vapor-HF (\mbox{V-HF}).}
    \textbf{a)} Top-view schematics of the \mbox{V-HF} lift-off process steps through the holes:
    1) The unpatterned \ce{Si3N4} mask layer.
    2) Integration of a hexagonal grid of holes in the stencil mask with a grid spacing of $d_{\mathrm{grid}}$ and $\diameter_{\mathrm{hole}}$. The hole size is chosen to block any deposition through them (see Fig.~\ref{figA1}).
    3) The \ce{SiO2} is etched via A-HF which creates an under-etch region around the holes with a radius of $(UE_{\mathrm{A-HF}})$. 
    4) The samples are then sent for material evaporation (yellow). Under the high deposition-angle, the holes will block the material from reaching the substrate (blue). 
    5) During the \mbox{V-HF} step, the \ce{SiO2} gets etched isotropically where the remaining etching distance is defined as: $d_{\mathrm{grid}} - 2 \times (UE_{\mathrm{A-HF}})$.
    6) This will enlarge the under-etch region until their collision with each other; forming a reversed hexagonal geometry. 
    7) Etching residues~\cite{Carter2000Surface} accumulate at the edges of the reversed hexagons: an etch front (red) is defined as where 2 opposite etching directions meet while an etch vertex (pink) is for 3 directions.
    8) These residues can be cleaned with a simple DI-water rinse, followed by IPA.
    \textbf{b-c)} Energy dispersive X-ray (EDX) spectrum and atomic force microscopy (AFM) images of before (step 7) and after (step 8) cleaning of \mbox{V-HF} residues. The square box indicate the EDX scanning area. The accumulation of residues on the etch fronts and vertices reveal a Si-based compound which is then cleaned by a DI-water and IPA dip (red area). The Al and O background EDX peaks come from the sapphire substrate. No fluorine (F) signal was detected.
    }
\end{figure*}

The first and second Al electrodes are e-beam evaporated at room temperature (in a Plassys Bestek MEB550s) and at an angle of $\alpha = 70$\textdegree{}, with a on-chip target thicknesses of~\SI{20}{\nano\meter} and~\SI{30}{\nano\meter}, respectively. To account for the deposition angle, the nominal evaporation thicknesses are set to~\SI{20}{\nano\meter}/$\cos(\alpha) \approx$~\SI{60}{\nano\meter} and~\SI{30}{\nano\meter}/$\cos(\alpha) \approx$~\SI{90}{\nano\meter}. Their deposition rate is~\SI{0.1}{\nano\meter\per\second}. The AlO$_\mathrm{x}$ tunneling barrier is formed by static oxidization of the first layer at~\SI{15}{\milli\bar} for 6 min. Scanning and transmission electron microscopy (SEM and STEM) of the JJ trilayer are presented in Fig.~\ref{figA2}. Energy dispersive X-ray (EDX) measurement scans could not reveal any fluorine (F) contamination inside the junction.

\clearpage
\paragraph{\mbox{V-HF} lift-off}
The chemical reaction between silicon dioxide (\ce{SiO2}) and hydrogen fluoride (\ce{HF}) can be written as follows~\cite{Witvrouw00Comparison}:

\begin{equation}
\begin{aligned}
    \ce{SiO2_{(s)} + 6HF -> H2SiF6_{(aq)} + 2H2O_{(l)}} \\
    \ce{H2SiF6_{(aq)} <=> SiF4_{(g)} + 2HF}
\end{aligned}
\end{equation}
or simply, 
\begin{equation}
\begin{aligned}
    \ce{SiO2_{(s)} + 4HF}
    \ce{-> SiF4_{(ads)} + 2H2O_{(ads)}}
\end{aligned}
\end{equation}

In our case, we utilize vapor-phase HF (\mbox{V-HF}) to etch the SiO$_2$ and thereby lift-off the entire stencil mask, taking advantage of its selectivity against AlO$_\mathrm{x}$. In fact, this forms natively on the Al surface and acts as an effective etch-stop layer~\cite{Bakke05Etch, Saloniemi12Study}. As a result, the functional material of our Al-based transmon qubit is natively protected from the \mbox{V-HF}. Other materials may be affected differently after \mbox{V-HF} exposure~\cite{Bakke05Etch, guillemin2019etching}.

To ensure a reliable stencil mask lift-off and reasonable etching times, small holes are dotted in a hexagonal grid all-over the chip area. 
This grid is incorporated into the design and written in the same e-beam lithography step. Its size determines the hole density and etching time in \mbox{V-HF}. In various iterations, we had a grid spacing ranging between $\sim$ 2--5~\SI{}{\micro\meter}. The holes keep a safe distance, less or equal to the grid spacing, from all other edges of the design elements.
The hole diameter ($\diameter_{\mathrm{hole}}$), along with the deposition angle $\alpha$ and the thickness of the top mask layer $th_{\mathrm{Si_{3}N_{4}}}$, are chosen to block material deposition onto the substrate (check paragraph \textit{a} and Fig.~\ref{figA1}b-c) according to:
\begin{equation}
{max(\diameter_{\mathrm{hole}})} = (th_{\mathrm{Si_{3}N_{4}}}) \cdot \tan\left(\alpha \cdot \frac{\pi}{180}\right)
\end{equation}

Practically, we also have to consider the sidewall deposition thicknesses of the intended electrodes in order not to clog the holes. Taking all of this into account, we opted for $\diameter_{\mathrm{hole}} \sim$~\SI{150}{\nano\meter}, $\alpha =$ 70\textdegree{} for $th_{\mathrm{Si_{3}N_{4}}} =$~\SI{100}{\nano\meter}. The high deposition angle avoids having stray fingers near the JJ in the Manhattan-style configuration, see Fig.~\ref{fig2}b. 

To complete the lift-off we expose our samples to vapor-HF using a Primaxx$^\text{\textregistered}$ uEtch tool~\cite{uEtch_SPTS_KLA}. We use a standard recipe with the following set parameters: Pressure (125 torr), HF (310 sccm), Alcohol (EtOH, 350) and N$_2$ (1250 sccm). This specific process selectively etches the \ce{SiO2} according to:
\begin{equation}
\begin{aligned}
     \ce{SiO2_{(s)} + 2HF2^-_{(ads)} + 2AH^+_{(ads)}} \\
     \ce{-> SiF4_{(ads)}  + 2H2O_{(ads)} + 2A_{(ads)}}
\end{aligned}
\end{equation}
where A denotes the alcohol in use. The lateral etching distance is determined by calculating the grid spacing (hole-hole) minus twice the under-etch distance from the A-HF step: $d_{\mathrm{grid}} - 2 \times (UE_{\mathrm{A-HF}})$. 
The vapor-HF gas isotropically etches away the sacrificial \ce{SiO2} layer and therefore detaches the rest of the stencil mask from the substrate. When two (or three) etching direction meet, we get an accumulation of fluorine/silicone-based by-products~\cite{Carter2000Surface}, (i.e. \ce{SiF4} and \ce{H2SiF6}, some which are volatile), at the etch-fronts (or etch-vertices). It turns out, the leftovers on the sample are water-soluble and cleaned away with a DI-water dip followed by a final IPA rinse.


\section{\label{App_Sim}Simulations}
\setcounter{figure}{0}  

\begin{figure*}[h]
    \includegraphics{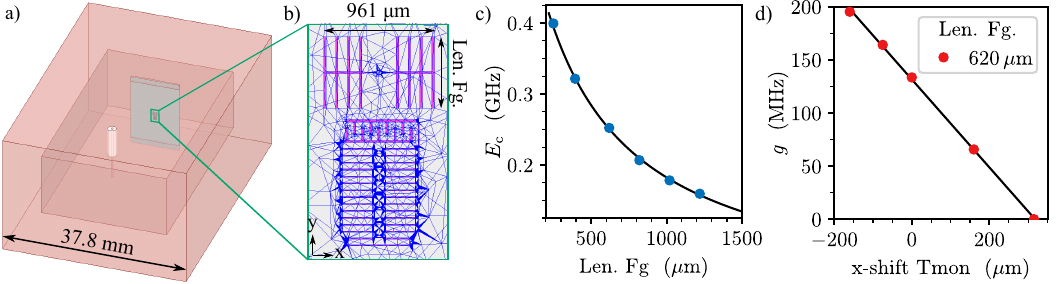}
    \caption{\label{fig_App__Ansys_Mesh}
    \textbf{Eigenmode finite element simulations with ANSYS.}
    \textbf{a)} Copper waveguide in which we perform the eigenmode simulations without qubit drive port. \textbf{b)} Mesh details (blue lines) of the Transmon-readout system (purple) in the center of the sapphire chip. The qubit and resonator are separated by 100~\SI{}{\micro\meter} along the y-axis. 
    \textbf{c)} We adapt the finger length to tune $E_\text{C}$ of the qubits, so that different ratios of $E_\text{J}/E_\text{C}$ can be accessed. \textbf{d)} By moving the qubit along the x-axis we vary the coupling strength $g$ between qubit and resonator as shown here for an example Transmon with a finger length of 620~\SI{}{\micro\meter}.
    }
\end{figure*}

To design the qubit-readout system we performed eigenmode simulations with the finite element solver ANSYS HFSS as shown in fig.~\ref{fig_App__Ansys_Mesh}. By varying the transmon capacitor fingers length and the lumped element junction capacitance and inductance different ratios of $E_\text{J}/E_\text{C}$ are accessible. We tune the dispersive shift $\chi$, that depends on the coupling strength $g$ between qubit and resonator, by shifting the qubit and resonator horizontally.

\section{\label{App_Meas}Measurements}
\setcounter{figure}{0}  

\paragraph{RT resistance}
2-point room temperature resistance measurements were carried on the test SQUID junctions ($\approx$100 units) on the same wafer as the qubit chips, with varying single junction size between 0.007 and~\SI{0.053}{\micro\meter\squared}. The measurements were done immediately after evaporation (performed at KIT) as well as before and after \mbox{V-HF} exposure (performed at FZJ \& TU Delft).
We record a junction yield of $\approx 95\%$ with their values summarized in Table~\ref{tab:junction_summary}. We attribute the changes in the calculated current densities to junction aging over time and to the effect of the \mbox{V-HF} processing, both of which could increase the room temperature resistance.

\begin{table}[h]
\centering
\begin{tabular}{lcc}

\textbf{Stage} & \textbf{$R_{RT}$ (k$\Omega$)} & \textbf{$\overline{J_c}$ (A/cm$^2$)} \\
\hline
After evaporation      & 5.6–45   & 28.2 \\
Before \mbox{V-HF}            & 6.9–50   & 19.3 \\
After \mbox{V-HF}             & 17.1–75  & 7.2 \\
\hline
\end{tabular}
\caption{Room-temperature resistance range and the average estimated critical current density at different process stages.}
\label{tab:junction_summary}
\end{table}

\begin{figure*}[h]
    \includegraphics{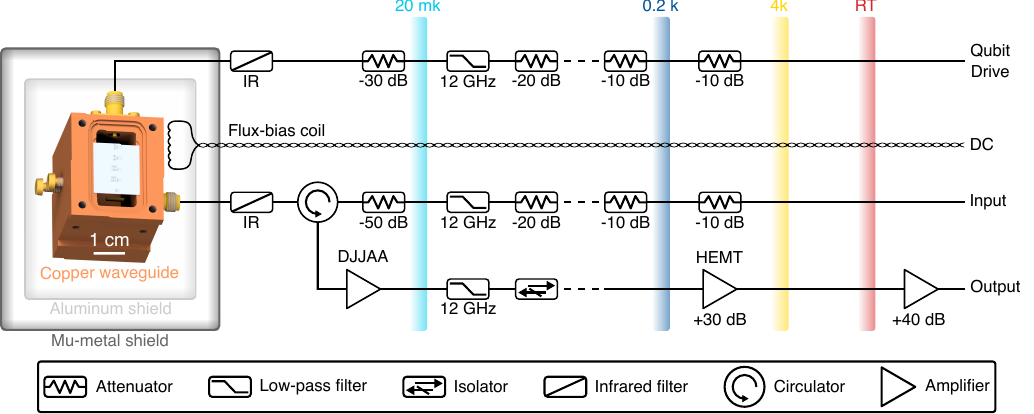}
    \caption{\label{figA7}
    \textbf{Copper waveguide \& Cryogenic setup.} 
    The qubit chip is loaded in a 3-dimensional copper waveguide, which is then covered in an aluminum and mu-metal shield and placed at the mixing chamber of the cryostat. The waveguide has two ports with non-magnetic pins. One of them connects to a drive line and the other is connected to the input/output of the qubit through a circulator. The output signal first passes through a dimer Josephson-junction-array amplifier (DJJAA) before going the rest of the components in the setup. A flux-bias coil wraps around the cover of the waveguide (not drawn for clarity) and is connected to a DC bias source.
    }
\end{figure*}

\paragraph{Waveguide and cQED Setup}
After fabrication, the stencil qubit chip is mounted in a copper waveguide for circuit quantum electrodynamics (c-QED) measurements. The waveguide is place inside a Al and a mu-metal magnetic shields. The sample is cooled down to a base temperature of \SI{20}{\milli\kelvin}. A schematic of the cryogenic setup is presented in~\ref{figA7}.

\paragraph{Further Qubit Data}

\begin{figure*}[h]
    \includegraphics{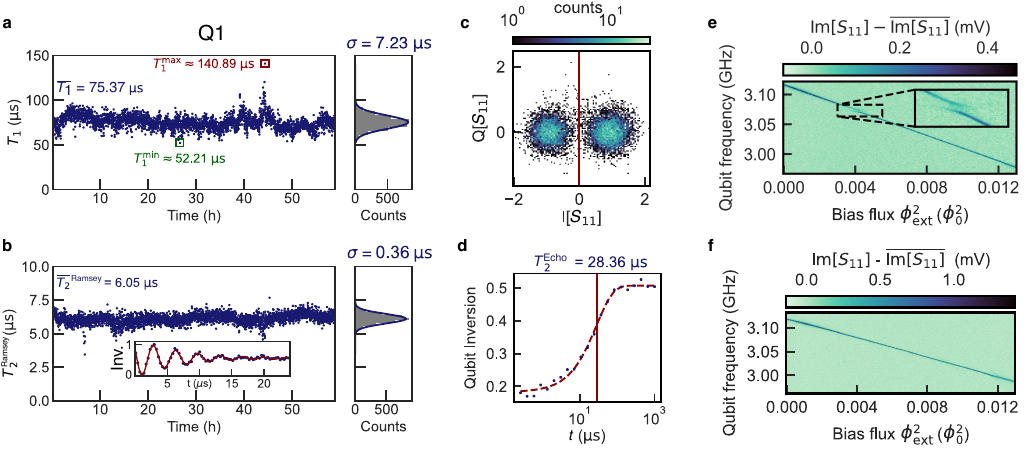
    }
    \caption{\label{figA4}
    \textbf{Further qubit Q1 measurements}
    \textbf{a-b)} Statistical measurement of $T_1$ and $T_2$ over a 60 hour period. The lower $T_2$ compared to the value reported in the main text can be attributed to the setup changes we made; i.e. by adding more capacitors on the drive line that isolates the qubit from charge noise, the $T_2$ was increased as shown in the main text.
    \textbf{c)} Single shot readout clouds after a $\pi/2$-pulse.
    \textbf{d)} $T_2^\mathrm{Echo}$ = \SI{28.36}{\micro\second}.
    \textbf{e-f)} Two-tone spectral purity of the qubit frequency against applied flux. Panel e) shows an avoided crossing (inset), which after a thermal reset, vanishes in panel f).
    }
\end{figure*}

\begin{figure*}[h]
    \includegraphics{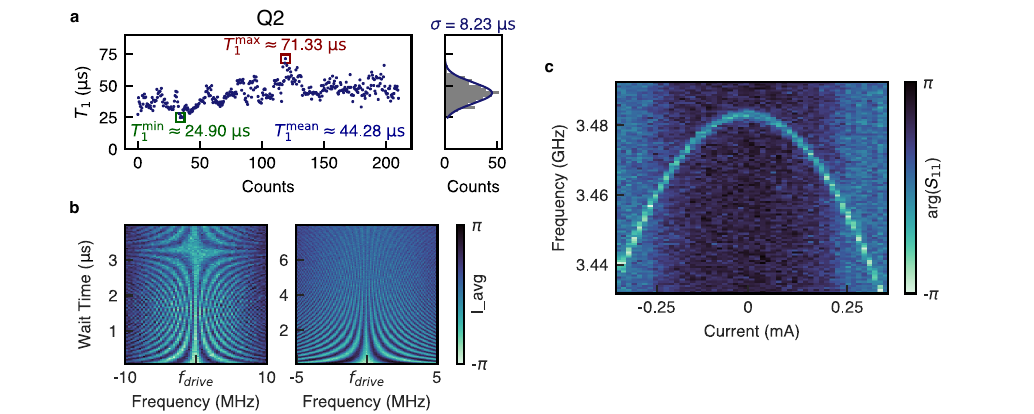
    }
    \caption{\label{figA6}
    \textbf{Further qubit Q2 measurements.}
    \textbf{a)} Statistical measurement of $T_1$ repeated over 200 times.
    \textbf{b)} Ramsey fringes at the drive frequency where the charge offset was not at a degeneracy point (left) and where it was (right).
    \textbf{c)} Two-tone flux spectroscopy of qubit Q2.
    }
\end{figure*}

Table~\ref{tab:T1_summary} provides the line-cuts value of Q1 from Fig.\ref{fig5} in the main text. Further measurements on Q1 and Q2 are shown in Fig.~\ref{figA4} and ~\ref{figA6}.

\begin{table}[!h]
\centering
\begin{tabular}{lccc}

\textbf{Type} & \textbf{Point} & \textbf{$\bar{T}_1$ (\SI{}{\micro\second})} & \textbf{$\sigma$ (\SI{}{\micro\second})} \\
\hline
\multirow{3}{*}{Time cut} 
  & $t = 0.28$ h  & 54.36 & 17.16 \\
  & $t = 3.84$ h  & 54.01 & 15.46 \\
  & $t = 6.69$ h  & 55.60 & 14.87 \\
\hline
\multirow{3}{*}{Flux cut} 
  & $\Delta f_q = 25.05$ MHz & 38.70 & 9.77 \\
  & $\Delta f_q = 87.01$ MHz & 69.10 & 6.12 \\
  & $\Delta f_q = 94.74$ MHz & 97.40 & 7.97 \\
\hline
\end{tabular}
\caption{Mean and standard deviation of $T_1$ extracted from Fig.~\ref{fig5}b–c for time (blue) and flux (red) cuts.}
\label{tab:T1_summary}
\end{table}

\clearpage

\twocolumngrid

\bibliography{Ref} 

\end{document}